\documentclass[%
reprint,
 amsmath,amssymb,
aps,
prl,
notitlepage,
]{revtex4-2}
\usepackage{graphicx}% Include figure files
\usepackage{dcolumn}% Align table columns on decimal point
\usepackage{bm}% bold math
\usepackage{hyperref}% add hypertext capabilities
\usepackage{textgreek}
\usepackage{xcolor}
\usepackage[caption=false]{subfig}%labelled subfigures

\begin{document}
\preprint{APS/123-QED}

%\title{Switching fronts, dark solitons and nonlinear oscillations for counterpropagation \\
%in ring resonators with normal dispersion}
\title{Frequency Comb Enhancement via the Self-Crystallization of Vectorial Cavity Solitons}

\author{Graeme N. \surname{Campbell$^{1,2}$}}
\email{graeme.campbell.2019@uni.strath.ac.uk}
\author{Lewis \surname{Hill$^{2}$}}
\author{Pascal \surname{Del'Haye$^{2,3}$}}
\author{Gian-Luca \surname{Oppo$^{1}$}}
\affiliation{$^1$SUPA and Department of Physics, University of Strathclyde, Glasgow, G4 0NG, Scotland, UK\\ %107 Rottenrow, 
$^2$Max Planck Institute for the Science of Light, 91058 Erlangen, Germany\\ $^3$Department of Physics, Friedrich-Alexander-Universit\"at Erlangen-N\"urnberg, 91058 Erlangen, Germany
}

\begin{abstract}
    Long range interactions between dark vectorial temporal cavity solitons are induced though the spontaneous symmetry breaking of orthogonally polarized fields in ring resonators. Turing patterns of alternating polarizations form between adjacent solitons, pushing them apart so that a random distribution of solitons along the cavity length reaches equal equilibrium distances. Enhancement of the frequency comb is achieved through the spontaneous formation of regularly spaced soliton crystals, `self-crystallization', with greater power and spacing of the spectral lines for increasing soliton numbers.  
\end{abstract}

%\begin{keyword}
%Micro-Resonator \sep Counter-Propagation \sep Normal Dispersion \sep Nonlocal \sep Lugiato-Lefever Equation \sep Domain Wall \sep Dissipative soliton 
%\end{keyword}

\maketitle

%\tableofcontents

%main content

%\section{Introduction}\label{sec:intro}

The generation of optical frequency combs \cite{PasquaziReview18} is an active area of research due to the wide range of practical applications that span across various fields including telecommunication \cite{pfeifle2014coherent,pfeifle2015optimally}, spectroscopy \cite{suh2016microresonator,dutt2018chip} and quantum technologies \cite{reimer2016generation}. Temporal cavity solitons (TCS) \cite{coen2016temporal} can be key
elements for broadband optical frequency combs \cite{kippenberg2011microresonator}. TCS are a special class of cavity solitons that originate in dissipative optical resonators under the action of external driving, diffraction \cite{scroggie1994pattern,firth1996optical} and/or group velocity dispersion. Ring resonator geometries are now regularly used for the generation of optical frequency combs via TCS \cite{PasquaziReview18}.

\begin{figure}[]
    \centering\includegraphics[width=1\linewidth]{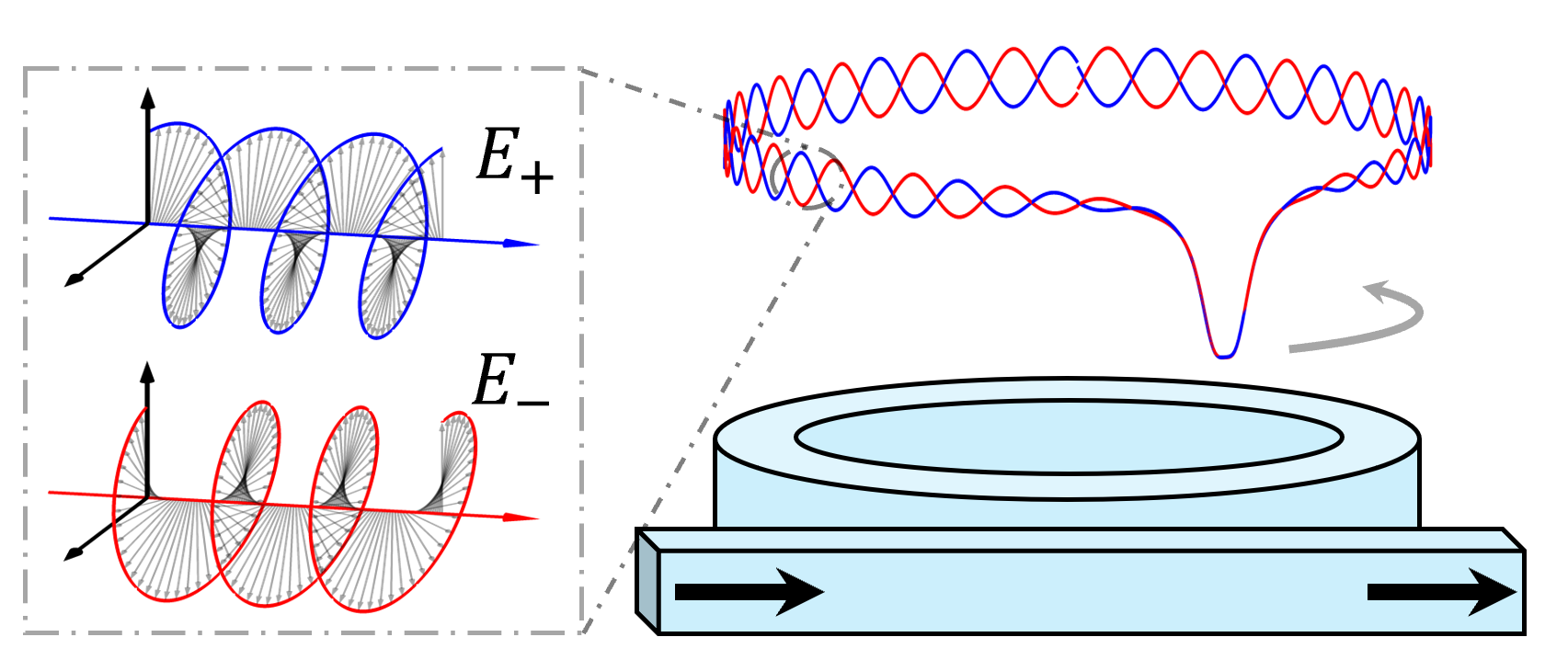}
    \caption{A ring resonator composed of a Kerr nonlinear medium. Linearly polarised light is coupled in and out of the resonator via a waveguide. An example intracavity power profile of a vectorial soliton is shown, presenting a Turing pattern of broken symmetry between fields of opposite circular polarization, visible as out-of-phase oscillations in the background of the dark soliton pulse.}
    \label{fig:setup}
\end{figure}

We consider a high finesse ring resonator composed of a Kerr medium, see Fig. \ref{fig:setup}, in the normal dispersion regime. A linearly polarized driving laser is coupled into the cavity, such that the intracavity fields may be resolved into components of orthogonal polarizations. In considering polarization components, vectorial TCSs display features in addition to those seen for a cavity with a single field, due to the possibility of spontaneous symmetry breaking (SSB) between polarization components \cite{christodoulides1988vector}. The SSB of light within Kerr resonators has been demonstrated theoretically and experimentally where the intracavity field is composed of orthogonal polarized components \cite{garbin2020asymmetric,xu2021spontaneous,xu2022breathing,moroney2022Kerr,quinn2023random,coen2023nonlinear,quinn2023towards,Fatome2023eqec,Huang2024}, counterpropagaing components \cite{kaplan1981enhancement,kaplan1982directionally,wright1985theory,del2017symmetry,woodley2018universal,hill2020effects,woodleyPRL21,cui2022control,bitha2023complex}, a combination of the two \cite{Campbell2023Fabryperot2pol,hill2023symmetry,Hill2023multi}, and most recently, between two, or more, coupled resonators \cite{rah2024demonstration,Ghosh2023twinresonator,Cheah2023Spontaneous, ghosh2024controlled}. 

We investigate the polarization properties of vectorial dark cavity solitons (VDS) in the normal dispersion regime and its effects on the formation of frequency combs. In particular, we present a useful `self-crystallization' phenomenon in which an initially random distribution of VDSs spontaneously form a regular soliton crystal (RSC). Previously, the generation of RSCs has been demonstrated through perturbations introduced near avoided mode crossings \cite{cole2017soliton}, or an external modulation \cite{lu2021synthesized} of the field. Here we instead present long range interactions between adjacent VDSs via a SSB of Turing patterns described by the coupled Lugiato-Lefever equations (LLE) \cite{lugiato1987spatial,haelterman1992dissipative,lugiato2018lugiato,xu2021spontaneous,garbin2020asymmetric,hill2020effects,Geddes1994Polarisationpatterns}
\begin{align}
    \partial_t E_\pm = S - (1 + i\theta)E_\pm + i(|E_\pm|^2 + 2|E_\mp|^2)E_\pm - i\partial^2_\tau E_\pm,\label{eq:coupledLLEs}
\end{align}
where $E_\pm(\tau,t)$ are the slowly varying amplitudes of the two orthogonal polarization components of the field, $S$ is the amplitude of the input field, considered to be real and positive, and $\theta$ is the input pump detuning to the near nearest cavity resonance. $t$ is the `slow time' temporal variable describing the evolution over many round trips of the cavity, while $\tau$ is the `fast time' longitudinal variable describing the evolution over a single round trip of the cavity in the normal dispersion case with $0\leq \tau \leq \tau_\text{R}$, where $\tau_\text{R}$ is the resonator round trip. Eqs. (\ref{eq:coupledLLEs}) are invariant under the exchange of the $+$ and $-$ indices, the fundamental symmetry of the system. Stationary solutions satisfying $E_+ = E_-$ are symmetric and $E_+ \neq E_-$ are symmetry broken.

\begin{figure}
    \centering\includegraphics[width=1\linewidth]{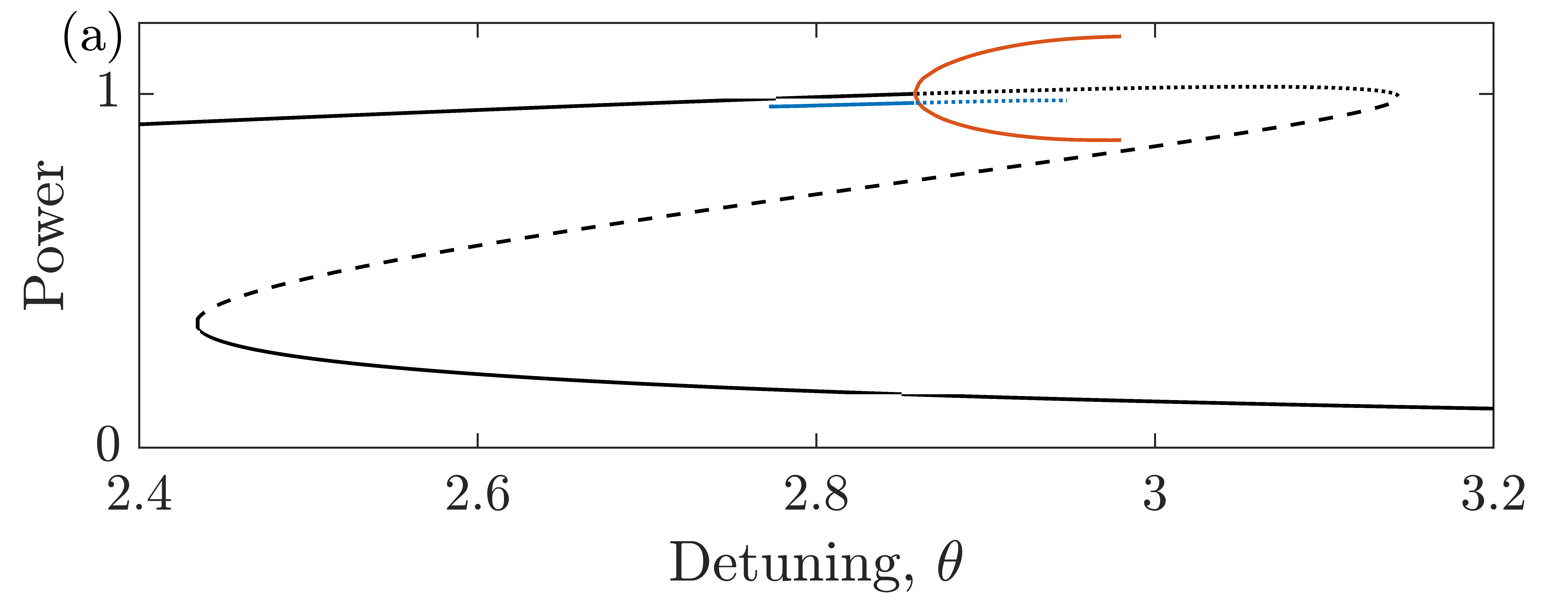}

    \includegraphics[width=1\linewidth]{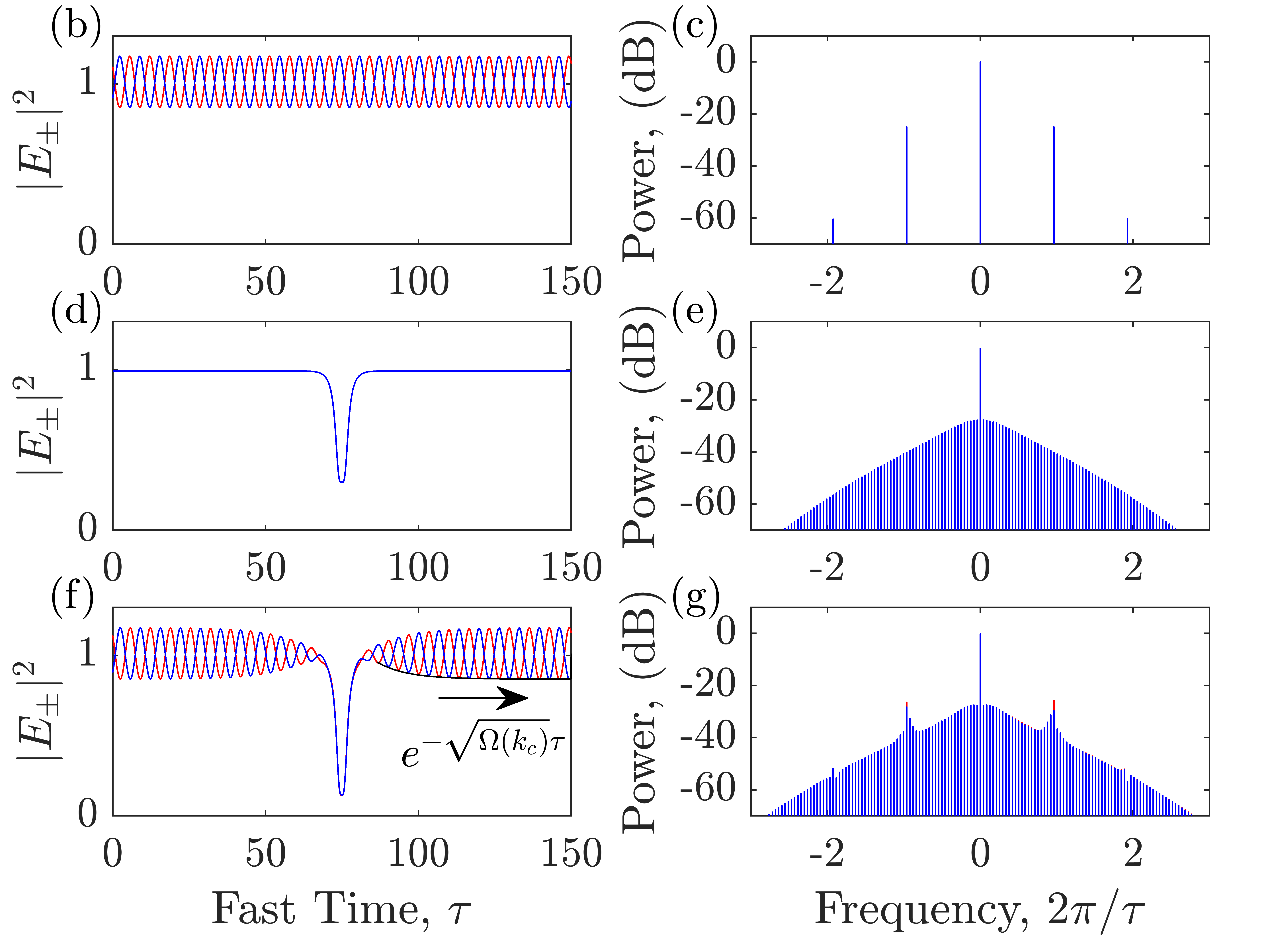}
    \caption{Solutions of Eqs. (\ref{eq:coupledLLEs}) for $S=1.01,\tau_\text{R}=150$. (a) Stable (solid black curves) and unstable (broken black curves) symmetric homogeneous solutions, and stable (solid blue curves) and unstable (broken blue curves) symmetric single dark soliton solutions plotted as their average power. The maximum and minimum power of stable symmetry broken Turing pattern are also shown, in red. (b) Turing pattern of alternating polarization for $\theta = 2.94$ and (c) the corresponding frequency comb. (d) Power profile of symmetric dark soliton solutions for $\theta = 2.8$ and (e) the corresponding frequency comb. (f) Power profile of symmetry broken dark soliton solutions for $\theta = 2.94$ and (g) the corresponding frequency comb. The black curve in (f) outlines the envelope of the Turing pattern $\propto \exp{(-\sqrt{\Omega(k_c)}\tau)}$ as it approaches the vectorial dark soliton.}
    \label{fig:HSSSSB}
\end{figure}

We first provide a description of the SSB of the homogeneous stationary solutions (HSS) of Eqs. (\ref{eq:coupledLLEs}). The HSS of Eqs. (\ref{eq:coupledLLEs}) correspond to the two coupled equations
\begin{equation}
    S^2 = H_\pm^2 - 2(\theta - 2H_\mp)H_\pm^2 + ((\theta - 2H_\mp)^2 +1 )H_\pm,\label{eq:HSS}
\end{equation}
where $H_\pm = |E_{0,\pm}|^2$ is the power of the HSS $E_{0,\pm}$. In Fig. \ref{fig:HSSSSB}a we plot solutions of Eq. (\ref{eq:HSS}) for $S=1.01, \tau_\text{R}=150$. For this value of $S$ there are only symmetric HSS ($H_+ = H_-$) which are plotted as the black curve. In the parameter region of our interest there are no symmetry broken HSS solutions. The symmetric HSS form a tilted Lorentzian curve, where stable solutions are plotted with solid lines and and unstable solutions as broken lines. We note that Eqs. (\ref{eq:coupledLLEs}) have undergone extensive investigation in the absence of fast time effects \cite{woodley2018universal,garbin2020asymmetric,hill2020effects,campbell2022counterpropagating}.

Of key importance here is a Turing instability due to a SSB bifurcation of the high power bistable symmetric HSS typical of regimes of normal dispersion, resulting in the formation of a Turing pattern stationary state formed of alternating orthogonal polarizations. This supercritical bifurcation occurs when increasing the detuning and is plotted as a red curve depicting the maximum and minimum powers of the Turing pattern in Fig. \ref{fig:HSSSSB}a. This instability is due to the field interaction through the cross-Kerr modulation and so it is not present on the high power HSS of a single LLE \cite{parra2016dark,godey2014stability}. The Turing instability is found by considering perturbations on the HSS of the form $E_\pm = E_{0,\pm} + \epsilon_\pm e^{ik\tau + \Omega t}$ where $k$ is the wavenumber of the perturbation and $\Omega$ is the slow time eigenvalue. The growth rate of this perturbation is then
\begin{align}
    \Omega(k) &= -1 \pm \sqrt{\frac{-A_1B_1 -A_2B_2 \pm Q}{2}},\label{eq:eigen}\\
    Q &= \sqrt{(A_1B_1 - A_2B_2)^2 + 4A_1A_2C^2},
\end{align}
where $A_{1,2} = \theta - k^2 - H_\pm - 2H_\mp$, $B_{1,2} = \theta - k^2 - 3H_\pm - 2H_\mp$, $C^2 = 8H_+H_-$. These eigenvalues have a similar form to the linear stability analysis of Refs. \cite{hill2020effects,woodley2018universal} where dispersion is neglected ($k=0$). From these eigenvalues we may approximate the Turing wavenumber from the critical wavenumber with largest growth, $\Omega(k_c)$. For the example Turing pattern shown in Fig. \ref{fig:HSSSSB}b we find a good agreement between the predicted $k_c \approx 0.96$ and measured $k\approx 1.01$ wavenumber, although the value of $\theta$ is well above the Turing instability threshold.

In the normal dispersion regime, Eqs. (\ref{eq:coupledLLEs}) exhibits VDSs \cite{parra2016dark}. These solutions are composed of localized switching fronts which connect the high and low power stable HSSs. Oppositely oriented pairs of switching fronts can `lock' due to the interaction of local fast time oscillations close to the lower power plateau and become stationary VDS. This mechanism of soliton formation was first proposed for spatial solitons composed of diffractive switching fronts \cite{rosanov1990diffractive,oppo1999domain,oppo2001characterization}, but has been demonstrated longitudinally in the ring resonator with a single field component theoretically \cite{parra2016dark} and experimentally \cite{garbin2017experimental}, as well as in Fabry-P\'erot configurations \cite{campbell2023dark}. 

First considering symmetric solitons, we note that at symmetry $E_+ = E_- = E$, Eqs. (\ref{eq:coupledLLEs}) reduce to
\begin{equation}
    \partial_\tau E = S - (1+i\theta)E + 3i|E|^2E -i\partial_\tau^2E.\label{eq:equalfield}
\end{equation}
This means that under a re-normalization of fields $E\rightarrow E/\sqrt{3}$, $S\rightarrow S/\sqrt{3}$ the stationary VDS of our system are analogous to those of the LLE. A branch of symmetric solutions of Eqs. (\ref{eq:coupledLLEs}) containing a single VDS is shown in \ref{fig:HSSSSB}a as the blue curve (plotted as the average power over a round trip to separate it from the HSS). At this parameter value ($S=1.01$) symmetric VDSs are stable for values of detuning below the Turing instability, shown in Fig. \ref{fig:HSSSSB}d-\ref{fig:HSSSSB}e. As the detuning is increased, the VDS symmetric solution undergoes a SSB of the homogeneous background from which the soliton hangs. This SSB results in the formation of a Turing pattern of alternating polarization components and is phenomenologically identical to the SSB of the HSS in the absence of the VDS. 

The frequency comb of a symmetry broken VDS is shown in Fig. \ref{fig:HSSSSB}g. It maintains a similar spectral envelope to that of the single symmetric VDS (Fig. \ref{fig:HSSSSB}e) but it develops sidebands due to the periodic modulations at the tails. The sideband peaks are reminiscent of those generated by dispersive waves due to higher order dispersion \cite{fujii2020dispersion}. Here they are achieved with second order dispersion and the contribution of the Turing pattern modulation. The power and separation of these peaks correspond to the spectral lines of the frequency comb of the Turing pattern, Fig. \ref{fig:HSSSSB}c.

An important property of symmetry broken VDSs is that the amplitude of the Turing pattern envelope decays as $[\exp{(-\sqrt{\Omega(k_c)}\tau)}]$, with $\Omega(k_c)$ given by Eq. \ref{eq:eigen}, from the place where the VDS tails are close to the unstable symmetric HSS to a saturation value of the modulated intensity. The black line in Fig. \ref{fig:HSSSSB}f shows this exponential decay matching the Turing pattern minima at the tails of the VDS. We have verified that such agreement persists over a wide range of detunings and input pumps where symmetry broken VDS are found.

\begin{figure}
    \centering
    \includegraphics[width=1\linewidth]{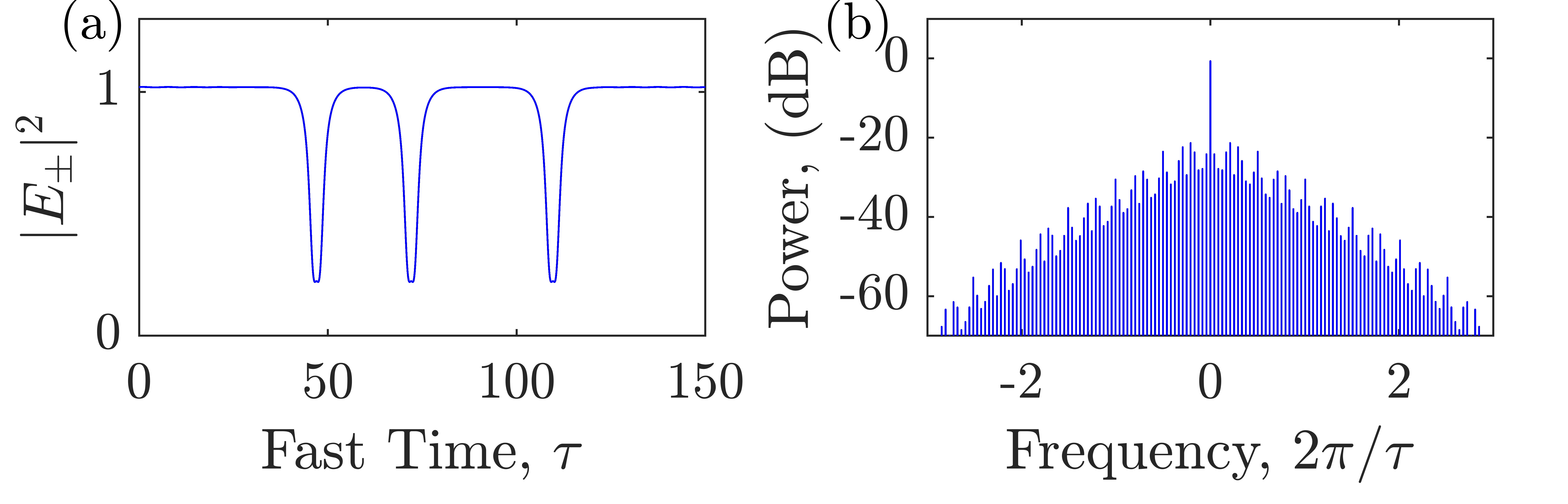}
    \includegraphics[width=1\linewidth]{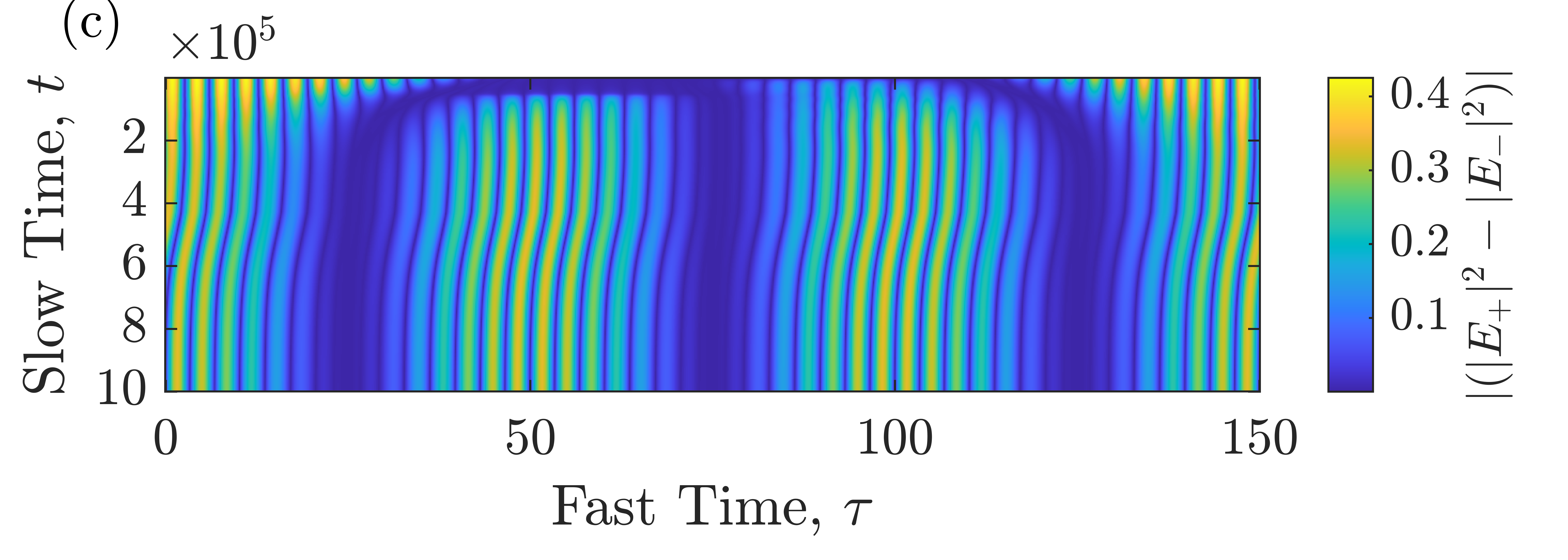}
    \includegraphics[width=1\linewidth]{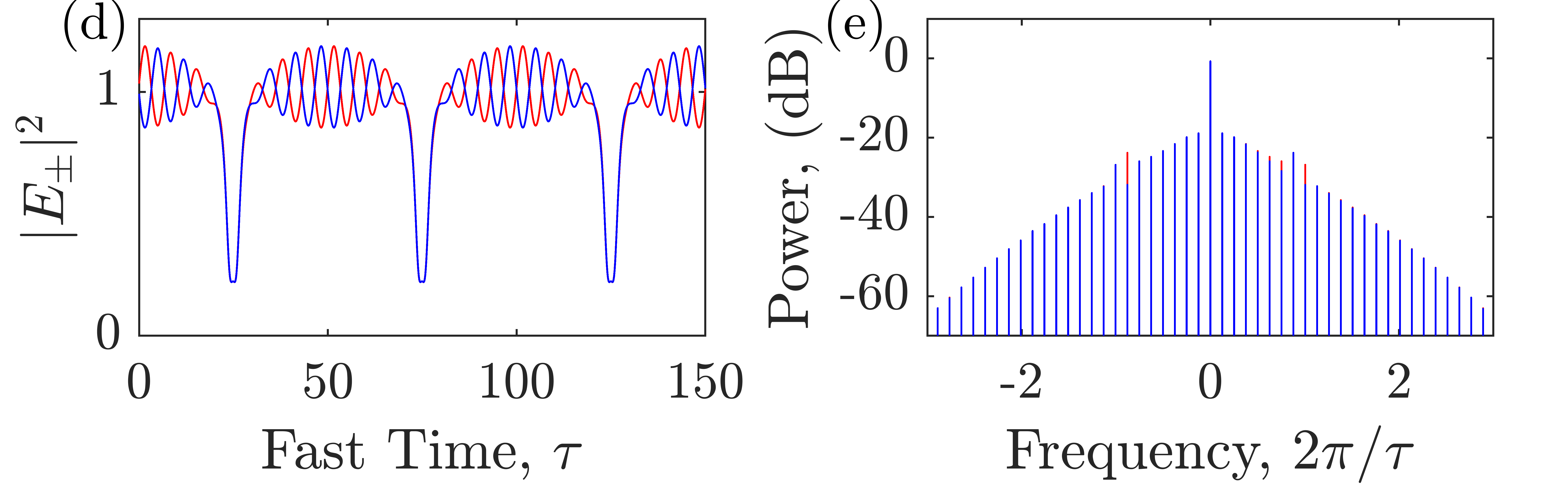}
    \caption{Evolution of three symmetry broken vectorial dark solitons for $S=1.02$, $\theta = 2.91$. Starting from an equal field condition (a), VDSs undergo SSB and move through the cavity as shown in (c), finally reaching a the RSC stationary state (d). Frequency combs corresponding to a random distribution of VDS (the initial condition shown in \ref{fig:darksolitonevolution}a) and a symmetry broken RSC (the final state shown in \ref{fig:darksolitonevolution}c) are presented in (b) and (e) respectively.}
    \label{fig:darksolitonevolution}
\end{figure}

We now consider solutions containing multiple VDSs along the cavity length simultaneously. After the SSB bifurcation, such solutions form Turing patterns in the intervals between VDSs. As the Turing patterns grow, adjacent VDS are `pushed' apart until an equilibrium of the pattern's amplitude is reached on both sides of the VDS, as shown in Fig. \ref{fig:darksolitonevolution}. The formation of the symmetry broken Turing pattern is hence found to introduce long range repulsive interaction between adjacent VDSs. Note that the symmetric VDSs do not exhibit these long range interactions and the VDSs remain stationary at arbitrary separation distances. 

\begin{figure}
    \centering
    \includegraphics[width=1\linewidth]{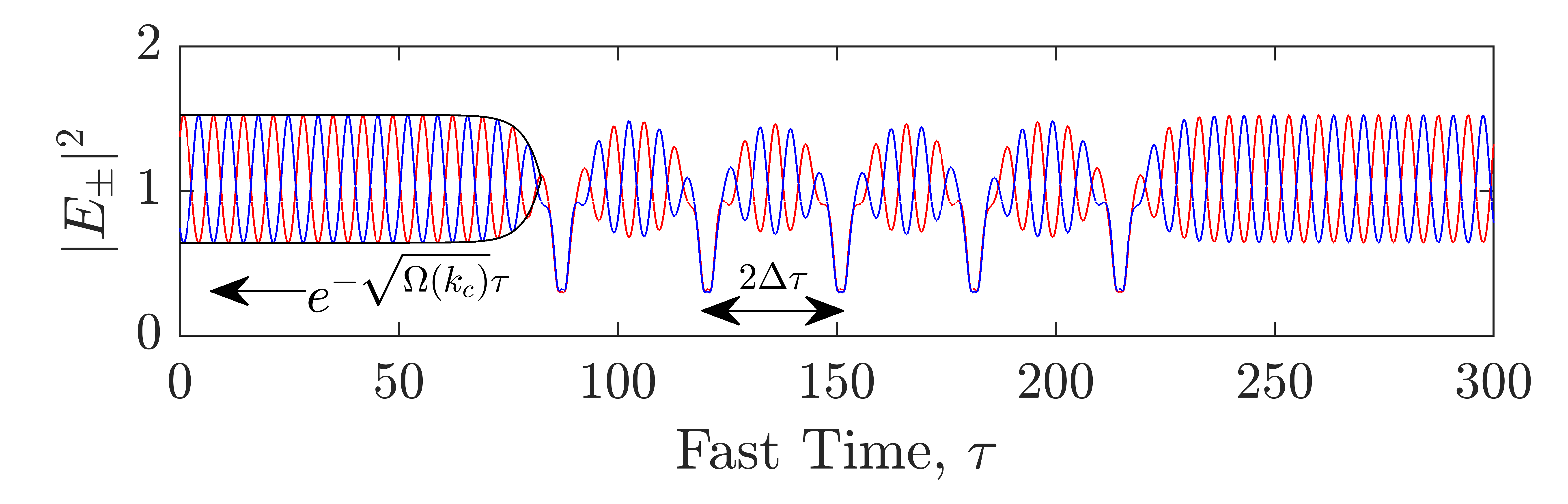}
    \caption{Partial soliton crystal composed of five VDSs for $S=1.06, \theta = 3$. The VDS can move no closer than $2\Delta\tau$ due to the repulsive interactions induced by the Turing pattern. The black curve follows the Turing pattern envelope starting from full saturation towards the VDS.}
    \label{fig:partcrystal}
\end{figure}

In Fig. \ref{fig:darksolitonevolution}a we start with three symmetric VDSs randomly distributed along the round trip for $S=1.02$, $\theta = 2.94$ and $\tau_\text{R} = 150$. For these parameter values, the homogeneous background is unstable to the formation of Turing pattern of alternating polarizations. The maximum amplitude reached by the Turing patterns in the intervals separating the VDSs depends on the separation of adjacent VCSs. As the pattern amplitude grows, the VDS are `pushed' along the resonator until an equilibrium configuration of the pattern is reached on either side of each VDS. The slow time evolution of the three VDSs is shown in Fig. \ref{fig:darksolitonevolution}c through direct numerical integration of Eqs. (\ref{eq:coupledLLEs}). Here it can be seen that the VDSs move such as to spread out along the cavity coordinate. This evolution ends in the stationary state shown in Fig. \ref{fig:darksolitonevolution}d composed of VDSs located equidistantly on the round trip of the cavity and separated by Turing patterns equal amplitude, thus forming a perfect soliton crystal.

The formation of such a RSC induced by SSB evolves spontaneously from the initial condition of three randomly positioned dark solitons. The organization process corresponds to self-crystallization from a random distribution of VDSs. The RSCs of our system are robust to a change in the number of VDS as the repulsive interaction will redistribute VDS to equidistant locations, as long as the new RSC spacing is shorter than twice the characteristic Turing patterns saturation length $\Delta\tau$, defined as the fast time distance where the pattern amplitude reaches its maximum value.

As can be seen in Fig. \ref{fig:darksolitonevolution}e, the RSC produces a frequency comb with a smooth spectral envelope and a spectral range three times larger than the frequency comb of the initial condition shown in Fig. \ref{fig:darksolitonevolution}b. In general, a RSC composed of $N$ VDSs produces a frequency comb equivalent to a single VDS in a cavity with round trip $\tau_\text{R}/N$. The RCSs emulate smaller cavity sizes, such that with increasing soliton number, a frequency comb with enhanced power and greater spacing of the spectral lines is obtained. Due to these features, the spontaneous formation of RSC has many potential applications, such as satellite communications \cite{federici2010review}, photonic radar \cite{riemensberger2020massively} and radio-frequency filters \cite{xu2018advanced,hu2020reconfigurable}. Being a self-organized structure, the RSC of our system offer different ways to generate and control RSCs than those demonstrated in Refs. \cite{cole2017soliton,lu2021synthesized}. %Additionally, the frequency combs produced in our system show regular peaks in the spectral envelope reminiscent of higher order dispersion.
As mentioned earlier, regular peaks in the spectral envelope are due to the Turing pattern wavenumber that is required for self-crystallization. Such peaks can be removed at will after self-crystallization by changing the control parameters across the SSB bifurcation, thus leaving a symmetric RSC with no pattern states between the VDS.

Even in the case of a small number of VDSs in a long cavity, such as in Fig. \ref{fig:partcrystal}, VDSs are found to move apart until a saturation of the Turing pattern amplitude is reached in the interval between them. In Fig. \ref{fig:partcrystal}, five VDSs have undergone SSB, and spread apart until the VDSs become stationary and produced a local RSC via self-crystallization. The maximum range of the repulsive interaction between VDSs can be investigated using the growth rate, Eq. \ref{eq:eigen}, of the critical wavenumber $k_c$ of the Turing pattern away from the VDS. We are able to estimate the a maximum interaction distance $2\Delta\tau \approx -2\ln(0.01|E_\text{max}|^2)/\sqrt{\Omega(k_c)}$, where we have assumed the VDS interaction disappears when the Turing amplitude reaches $1\%$ from the maximum amplitude $|E_\text{max}|^2$. This predicts a maximum lattice spacing of $2\Delta\tau \approx 32$ compared to the measured $2\Delta\tau \approx 30$ from Fig. \ref{fig:partcrystal}. The interaction distance of VDSs can then be controlled by changing the control parameters to alter the growth rate of the Turing patterns. A pair of VDSs will no longer interact should their separation be greater than $2\Delta\tau$ where the mediating Turing patterns reach saturation. By selecting a suitable cavity length and soliton number such that $N > \tau_\text{R}/2\Delta\tau$ we observe the self-crystallization phenomenon as is shown in Fig. \ref{fig:darksolitonevolution}. 

\begin{figure}
    \centering    
    \includegraphics[width=1\linewidth]{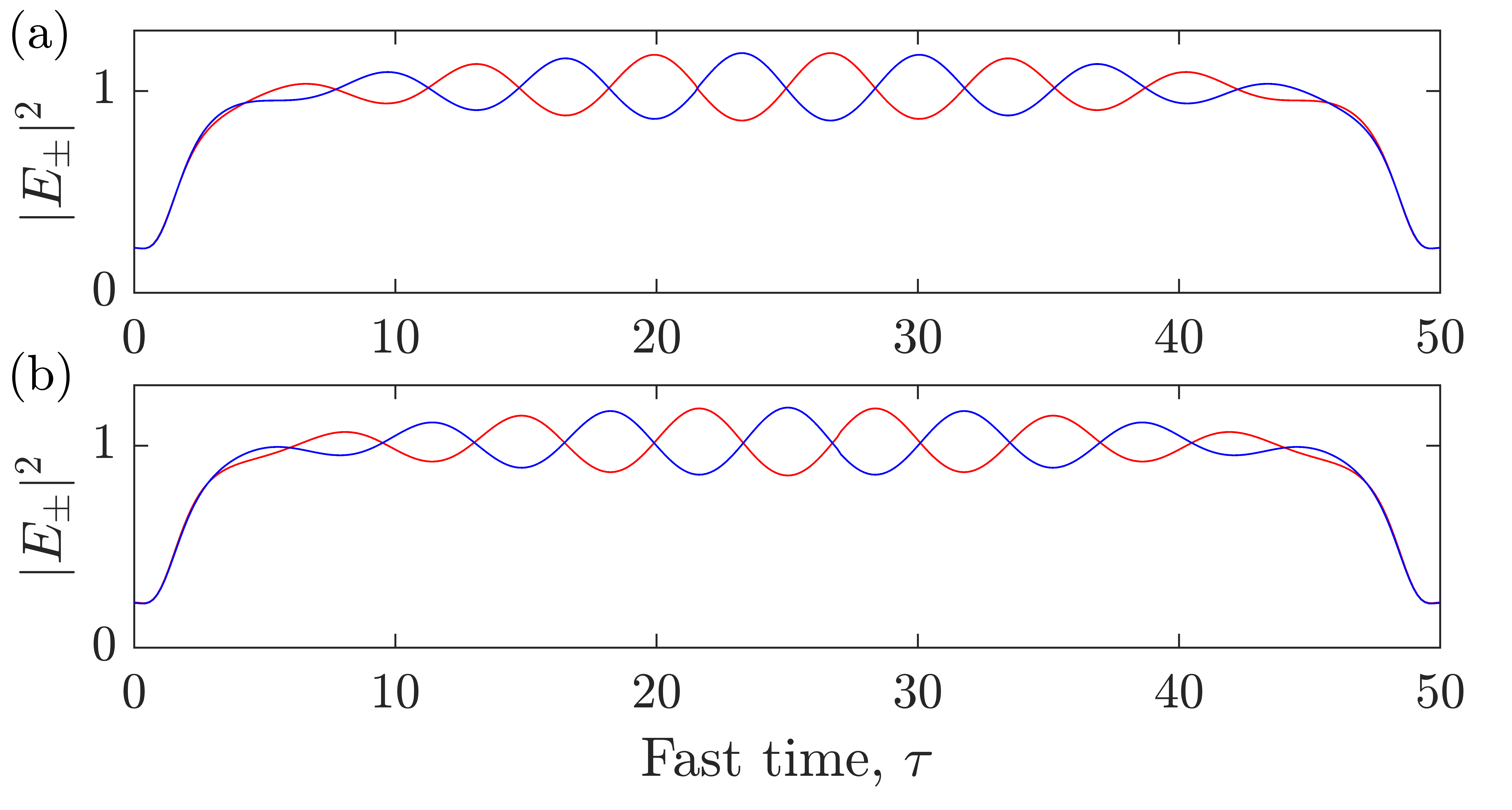}
    \caption{(a)-(b) Power profiles of RSC unit cells for $S = 1.02, \theta = 2.91, \tau_\text{R} = 50N$. Two additional unit cells can be obtained by exchanging the fields $E_+ \leftrightarrow E_-$ in both (a) and (b). The unit cells are related by a phase shift in the peaks of the Turing patterns of $\pi/2$.}
    \label{fig:defects}
\end{figure}

RSCs are composed of a unit cell which is perfectly repeating over the cavity round trip. The example RSC of Fig. \ref{fig:darksolitonevolution}d is composed of the unit cell shown in Fig. \ref{fig:defects}a repeated three times over the round trip. This solution is bistable with the RSC composed of the unit cell shown in Fig. \ref{fig:defects}b. These two unit cells possess the fast time symmetries $E_\pm(-\tau)= E_\mp(\tau)$ and $E_\pm(-\tau)= E_\pm(\tau)$ respectively, and two additional unit cells obtained by exchanging the fields $E_+\leftrightarrow E_-$ in Fig. \ref{fig:defects}. As such, there are four possible RSCs, each related by an integer multiple phase shift of $\pi/2$ in the peaks of the Turing pattern. We find that all four RSCs are stable and can be reached depending on the initial condition. 

If we return to Fig. \ref{fig:darksolitonevolution} we see that the evolution of the three VDSs is composed of two regimes. For slow time $t< 3\times 10^5$, the VDSs move apart due to the formation of the Turing patterns. At slow time $t\approx 3\times 10^5$, the three VDSs approach an equal spacing in the cavity, but here the Turing pattern displays a non integer $\pi/2$ phase shift with respect to the stationary unit cells presented in Fig. \ref{fig:defects}. We now see transient dynamics in which the equidistant VDSs lattice and Turing pattern drift in fast time at different rates. This drift continues until one of the four stationary configurations is reached. We note the possibility of forming `defective' crystals composed of alternating combinations of these four unit cells, which we leave for future publications.

In conclusion, the formation of a RSC is achieved from a random distribution of VDSs via pattern formation with two field components of orthogonal polarization. SSB results in the formation of Turing patterns of alternating polarization at the tails of the VDSs. Long range interactions between VDSs are induced and mediated by Turing patterns, which increase the separation between adjacent VCSs until an equidistant equilibrium distance is reached and a regularly spaced soliton crystal is formed. The growth of Turing patterns can be controlled by changing the control parameters, which determine the range of the interaction. Above the Turing instability, RSCs originate spontaneously (self-crystallization) without the need of any perturbation \cite{cole2017soliton,lu2021synthesized} and represents a new, readily implementable, method for RSC formation relevant for application \cite{federici2010review,riemensberger2020massively,xu2018advanced,hu2020reconfigurable}. RSCs produce a frequency comb displaying a smooth spectral profile and increased line spacing when compared to a random distribution of cavity solitons. %These desirable features are normally associated with a single VDS in a cavity of smaller round trip. 
As such, a RSC may be used to emulate smaller cavity sizes while avoiding the experimental limitations of small diameter ring resonators. The self-crystallization mechanism is universal in systems displaying temporal cavity soltions and Turing instabilities and has already been generalized to Fabry-P\'erot configurations with two orthogonal polarizations \cite{Campbell2023Fabryperot2pol}.

%%%%%%%%%%%%%

\vfill

We acknowledge support from the EPSRC DTA Grant No. EP/T517938/1. P.D. acknowledges support by the European Union's H2020 ERC Grant ``CounterLight'' 756966 and the Max Planck Society. LH acknowledges support from the SALTO scheme of the Max-Planck-Gesellschaft (MPG) and the CNRS.

\vfill

\bibliographystyle{unsrt}%ieeetran}
%\bibliography{MyBibFile.bib}

\end{document}